# XAO-assisted coronagraphy with SHARK-NIR: from simulations to laboratory tests

Umbriaco G.[a,b,c], Carolo E.[a,b], Vassallo D.[a,b,c], Farinato J.[a,b], Baudoz P.[d], Carlotti A.[e], Greggio D.[a,b], Marafatto L.[a,b], Bergomi M.[a,b], Viotto V.[a,b], Agapito G.[f,b], Biondi F.[a,b], Chinellato S.[g,b], De Pascale M.[a,b], Dima M.[a,b], D'Orazi V.[a,b], Esposito S.[f,b], Magrin D.[a,b], Mesa D.[a,b], Pedichini F.[h,b], Pinna E.[f,b], Portaluri E.[a,b], Puglisi A.[f,b], Ragazzoni R.[a,b], and Stangalini M.[h,b]

[a]INAF - Osservatorio Astronomico di Padova, Vicolo dell'Osservatorio 5, 35122, Padova, Italy
[b]ADONI - Laboratorio Nazionale Ottiche Adattive, National Laboratory for Adaptive Optics, Italy
[c]Dipartimento di Fisica e Astronomia, Università degli Studi di Padova, Vicolo dell'Osservatorio 3, 35122, Padova, Italy
[d]Lesia, Observatoire de Paris, PSL Research University, CNRS, Sorbonne Universites, Univ. Paris Diderot , UPMC Univ. Paris 06, Sorbonne Paris Cite, 5 place Jules Janssen , 92190 Meudon, France
[e]Institut de Planétologie et d'Astrophysique de Grenoble, 414, Rue de la Piscine, Domaine Universitaire, 38400 St-Martin d'Hères, France
[f]INAF - Osservatorio Astrofisico di Arcetri, Largo Enrico Fermi 5, 50125 Firenze, Italy
[g]INAF - Direzione Scientifica V.le del Parco Mellini 84, 00136 Roma, Italy
[h]INAF - Osservatorio Astronomico di Roma, Via Frascati 33, 00078 Monte Porzio Catone, Roma, Italy

## ABSTRACT

Several Extreme Adaptive Optics (XAO) systems dedicated to the detection and characterisation of the exoplanets are currently in operation for 8-10 meter class telescopes. Coronagraphs are commonly used in these facilities to reject the diffracted light of an observed star and enable direct imaging and spectroscopy of its circumstellar environment. SHARK-NIR is a coronagraphic camera that will be implemented at the Large Binocular Telescope (LBT). After an extensive simulation campaign, SHARK-NIR team selected a suite of coronagraphic techniques to be implemented in the instrument in order to fulfil the scientific requirements. In summary, the Gaussian Lyot coronagraph is the option to serve all those science cases requiring field-stabilization and moderate contrast. Observations in pupil-stabilized mode to search for exoplanets can take advantage of three Shaped Pupil masks (SPC ) and a Four-Quadrant Phase Mask (FQPM) coronagraph. The SPC are designed for high contrast on a small field close to the star and are robust to image and pupil jitter. The FQPM allows to access the entire scientific FoV (18"x18") and delivers excellent performance in ideal conditions (high Strehl ratios), but performance is still good, both close and further away from the star, even at lower Strehl and with moderate vibrations. After the procurement phase, the coronagraphic masks were delivered to our labs and we started to test their performance on the optical bench and define the alignment procedures that will be employed in the final integration of the instrument in our cleaning room. In this article, we describe the tests that we performed in the lab with SHARK-NIR coronagraphs. We measured the contrast achievable with each technique in very-high Strehl conditions and defined the alignment/integration procedures.

# 1. INTRODUCTION

SHARK is a coronagraphic camera proposed for LBT in the framework of the 2014 Call for Proposals for Instrument Upgrades and New Instruments.[1-4] We are building this tool because first of all we hold an excellent adaptive optic (AO) performance, we are in the northern hemisphere with a strong scientific case and the purpose is going on sky in a very short time. In order to take advantage of these points we proposed a simple camera that will allow classical imaging, coronagraphic imaging, and coronagraphic low/medium resolution spectroscopy. SHARK-NIR together with the SHARK-VIS channel will cover a wide wavelength domain, going from $0.6\mu m$ to $1.7\mu m$ (Y to H band). SHARK-NIR will offer extreme AO direct imaging capability on a field of view (FoV) of about 18"x 18", and a coronagraphic long-slit spectroscopic at small (100) and medium (700) spectral resolution. The main science case of SHARK-NIR is searching for giant planets. To succeed, high contrast is necessary. We also emphasize that the LBT AO SOUL upgrade[5] will further improve the AO performance, making it possible to extend the exoplanet search to target fainter than normally achieved by other 8-m class telescopes, and opening in this way to other very interesting scientific scenarios, such as the characterization of AGN and Quasars, normally too faint to be observed, and increasing considerably the sample of disks and jets to be studied.

# 2. CORONAGRAPHY AND SHARK-NIR

We tested several coronagraphic masks. After an extensive simulation campaign,[6,7] SHARK-NIR team selected a suite of coronagraphic techniques. The Gaussian-Lyot Coronagraph is similar to the classical Lyot.[8] Instead of a hard edge mask, a Gaussian modulation is applied to the electric field in the focal plane (FP). A Lyot stop is then put in the subsequent pupil plane. This solution allows to achieve a better contrast at small angular separations with respect to the classical Lyot configuration. We also studied the performance of the Shaped Pupil Coronagraph (SPC);[9] classical SPC simply consists of a binary transmission pattern applied in the pupil plane to generate a high contrast region in the subsequent focal plane. If the plane in which the high contrast zone is generated is not the detector plane, as in the SHARK-NIR case, then a hard edge mask can be placed to remove all light falling out of this zone. Finally, a stop is placed in the subsequent pupil plane to remove residual light diffracted at the border of the pupil by the FP mask, as in the classical Lyot configuration (see Figure 1). The coronagraph can generate high contrast all around the star (360° extent) or in two regions of variable extent (asymmetric discovery space). In Figure 2 the apodizer masks, its FP masks, and the simulated PSFs are shown on the left.

On-line fast tip-tilt correction is foreseen in SHARK-NIR. This allowed us to evaluate focal plane coronagraphs requiring high PSF stability, such as the FQPM.[10] This coronagraph suppresses on-axis starlight by means of a phase mask inducing a phase shift on specific areas of the focal plane. In particular, the mask is arranged according to a four-quadrant pattern: two quadrants on one diagonal without phase shift and the two other quadrants providing a $\pi$ phase shift (see the image on the right in Figure 2). Provided that the image of the star is exactly located at the center of the mask, the four beams combine in a destructive way and the stellar light is mostly rejected outside of the pupil area. Then a Lyot Stop (LS) is placed in this exit pupil to remove the diffracted starlight.

## 2.1 Coronagraphy on test bench

The bench setup is illustrated in Figure 3. An iris located 50mm before the apodizer plane is illuminated with a collimated beam. Two lenses, L2 and L3, focus the beam on the position of the CORO wheel, where the coronagraphic focal plane masks are located. The F-number delivered by the two lenses is such that the physical size at 633nm of the focalized spot matched the one in the same plane in SHARK-NIR in the NIR ($1.6\mu m$, which is the working wavelength of the coronagraphs). The FP is finally reimaged on a camera using two lenses, L4 and L5. If L5 is removed from the optical path, an image of the apodizer (APO) is formed on the camera.

The optical bench for the coronagraphic masks tests and alignment included:

1. 20mW He-Ne laser (633nm)

2. ND filters

3. Plane mirror [FM1]

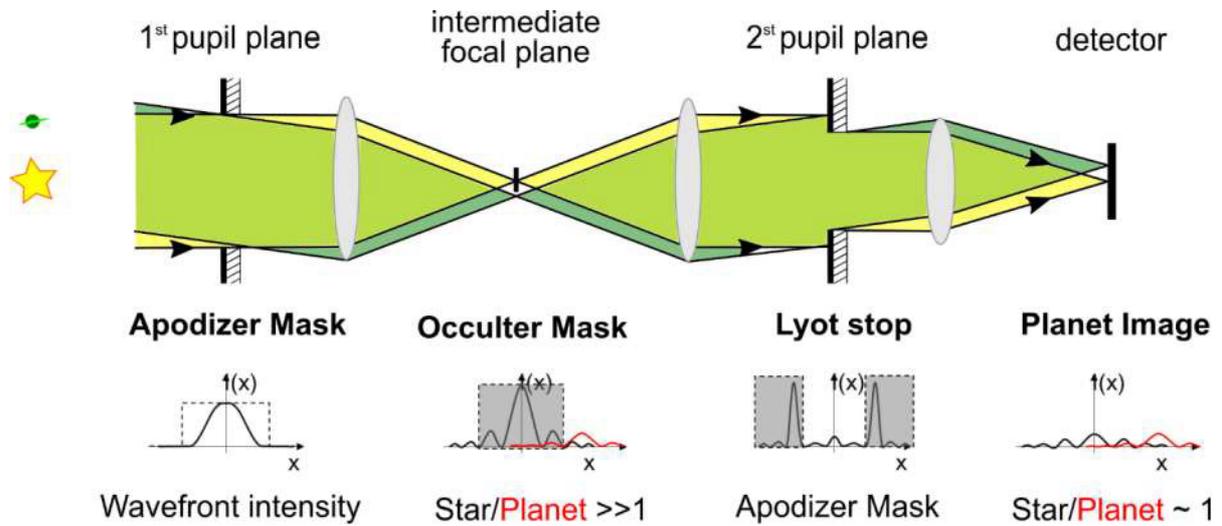

Figure 1. Coronagraphy technique in a scheme.

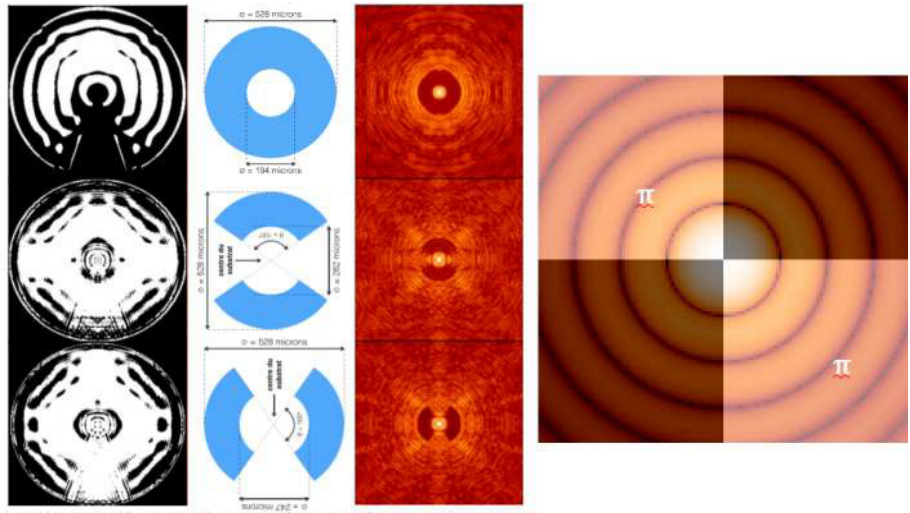

Figure 2. On the left: the Shaped Pupil coronagraphic masks. On the right: the Four Quadrant Phase Mask.

4. Spatial filter

5. Pellicle Beam Splitter [PBS]

6. Lens (BB2-E02 300mm 50.8) [L1]

7. Plane mirror [FM2]

8. Iris [Pupil]

9. Apodizer wheel [APO Wheel] on PI stage + Mercury controller

10. Lens (026-1730 D50.80 FL700 B MAR 400-700) [L2]

11. Plane mirror [FM3]

12. Lens (SINGLET 25.4X70mm actually 50mm) [L3]

13. Coronagraphic Focal Plane Mask (FPM) wheel [CORO wheel] on PI stage + Mercury controller

14. Lens (D50.80 F100mm P5630) [L4]

15. Lens ($\phi$=2" FL=150mm) [L5] (on linear stage PI M511.DG, pos. rep. 0.1$\mu$m)

16. Camera (AVT Pike F-145B pixel size 6.45$\mu$m resolution 16bit) [CCD]

In order to materialize a pupil plane after the FP mask, we introduced a deployable mirror (FM4) folding the light towards the center of the bench (see the green area in Figure 3). The beam folds after being collimated by L4, and the Lyot stops are located in the pupil plane. For instance, the stop is mandatory for FQPM to work, being the component that physically blocks the light. The upstream phase mask only diffracts it outside the pupil. On the other side, it is not a crucial component for the SPC. After the stop, a lens (L6) is placed in front of a CCD. Depending on the position of the CCD, L6 can form the image either of the source or of the pupil. Finally, by removing the folding mirror, it is possible to take images of the phase mask at higher resolution using an Allied Pike F-145 camera (CCD positioned at the focus of L5 with a sampling $\lambda/D$ with 10 px).

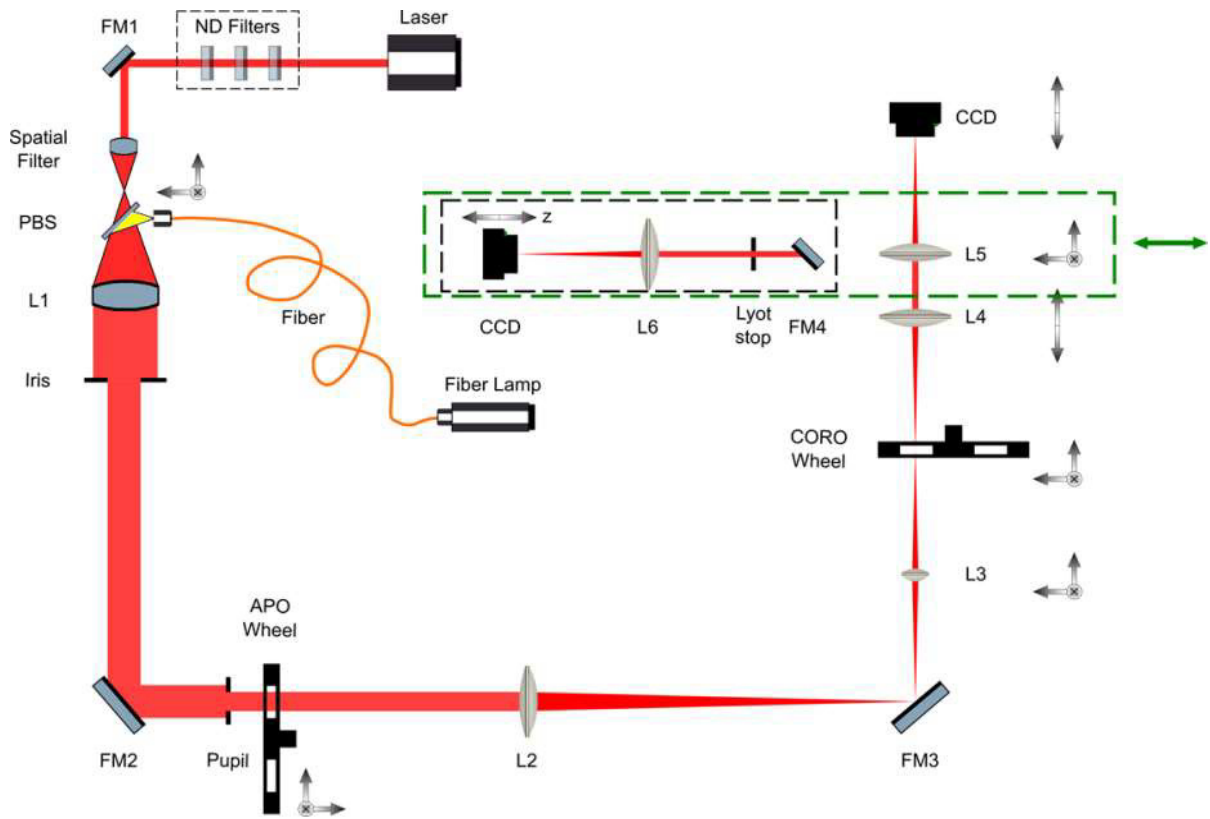

Figure 3. The bench optical scheme.

## 2.2 Masks alignment

### 2.2.1 Shaped Pupil masks

See Figure 3 for the optical scheme of the following procedure, the LS arm is not used.

1. Reference position [x0, y0] on the CCD is defined by taking an image of pupil (without L5 and without FPM) and its center is calculated by fitting a circle on it.

2. L5 and FPM are inserted. L5 is moved in z-direction to obtain the best focus of the FPM. The mask is illuminated with a diffuser to remove the laser coherence.

3. FPM is removed.

4. The PSF dimension ($32\mu$m) in the FPM plane is measured. FM3 is previously aligned in autocollimation with respect the incoming beam from L2. This ensure FM3 is used in cat's eye mode. We inserted a Foucault slit in the L2 focus (on the surface of FM3) and re-image the slit in FPM position by moving L3. To obtain this the diffuser is used between L2 and FM3, and 100 images to measure the sharpness of the Foucault slit are captured and averaged.

5. L5 on motorized and manual translation stage is moved for centering the PSF in [x0, y0].

6. L5 is removed.

7. An image of the pupil is captured to check the center of the pupil: [x0, y0]

8. L5 is inserted.

9. The PSF of the pupil is centered in [x0, y0] and its FWHM is calculated. In case the FWHM does not match the required size, L3 is moved and the best FP position after L3 is cheked by using the Foucault slit. Accordingly L4 is shifted of the same amount to ensure the beam is collimated. 100 images to measure the sharpness of the Foucault slit are captured and averaged.

10. APO is inserted, L5 is removed.

11. The APO best focus is found by moving the camera in z-direction. The diffuser is located between L2 and FM3.

12. APO is removed. The position of the center of the pupil [x0, y0] is checked.

13. L5 and FPM inserted.

14. L4 is moved by the same amount of the z-direction of the CCD.

15. L5 and FPM removed.

16. The pupil image with the apodizer inserted is checked. APO is centered by rotating the wheel, by radial movement and by moving in the x direction the stage. "Ad-hoc" python procedure are used to check the center.

17. L5 inserted, APO removed.

18. The PSF image is checked if it is in the center [x0,y0].

19. APO inserted.

20. FM2 is moved to shift the PSF center on the center of the optical axis [x0,y0], this is due to the wedged surface of the apodizers.

21. FPM inserted.

To calculate the raw contrast (see Section 2.3) a series of images for each SPC coronagraphic technique are captured:

1. Capture 1 image of SPC of apodizer PSF, low exposure time; the "off-axis" PSF.

2. Capture 1 image of SPC of apodizer + FPM, high exposure time; the coronagraphic PSF.

### 2.2.2 Four Quadrant Phase Mask

If we look at the ratio of the PSF peak intensity with and without the coronagraph (i.e. the attenuation) as a function of wavelength for a phase mask optimized to work at 1.6$\mu$m, we see a local minimum at a wavelength around 550nm (see Figure 4). For this reason, we decided to perform the tests and the alignment of the FQPM in the visible (which is much easier for several reasons), using a filter ($\lambda$=550nm, FWHM=40nm).

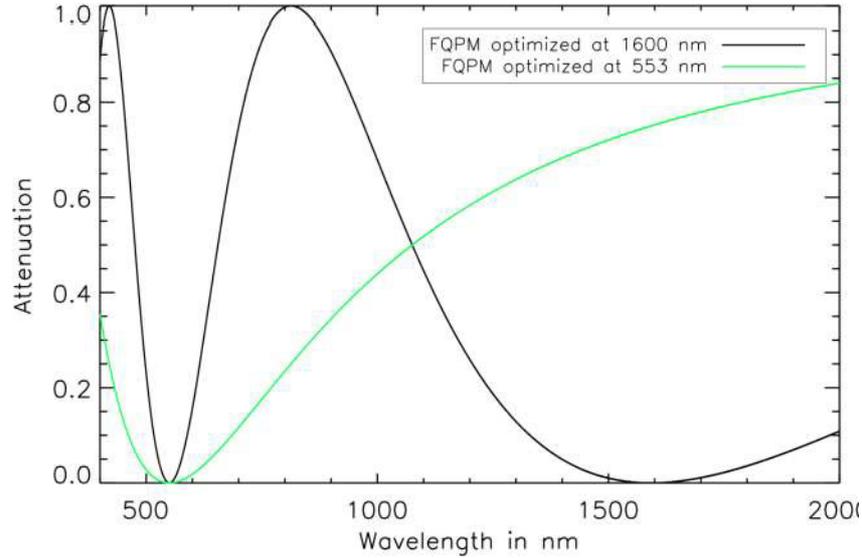

Figure 4. Comparison between theoretical PSF attenuation as a function of wavelength for a FQPM optimized at 1.6$\mu$m FQPM optimized at 553nm. The former can cancel a star at 550nm (sharp), but its high sensitivity to the bandwidth makes it worse in performance with respect to the latter at the end.

For the optical scheme of the following procedure see Figure 3. The LS arm is used in this case. Hereafter, CCD1 will refer to the camera positioned after L5, CCD2 to the camera in the LS arm.

1. Observing the PSF, the laser and the fiber spot of the Spatial Filter are co-aligned. After this, the laser is turned off.

2. The FQPM is inserted and we looked at the PSF on CCD1. The center of the mask is determined with a semiautomatic procedure that identifies the transitions between the quadrants given an image of the mask with diffused illumination of the FP.

3. The FQPM is removed and FM4 is inserted to fold the light towards the CCD2 arm.

4. CCD2 is moved on its linear stage to conjugate it to the pupil plane using a single lens (L6) relay near 1:1 magnification. We calculated the coordinates of the pupil center on CCD2 and its diameter with an "ad-hoc" Python script. The measured diameter is 1.916mm @633nm;

5. The LS is inserted and its position adjusted in z to focus it on the camera. Images of the LS are acquired and the coordinates of its center are calculated to properly align it in [x, y] to the reference position set by the pupil. The LS diameter is measured 1.498mm @633nm and 1.493mm @550nm, corresponding to around 80% of the pupil diameter, as desired.

6. Without moving CCD2 from its position, the FQPM is inserted. The characteristic diffraction pattern of the FQPM appears on the camera, made of 4 bright corners and an almost dark pupil (the coronagraphic pupil, hereafter, Figure 5).

7. Images of the coronagraphic (on-axis) PSF are captured, then the spot was off-centered by rotating the wheel and some off-axis PSFs are acquired (necessary to compute raw contrast).

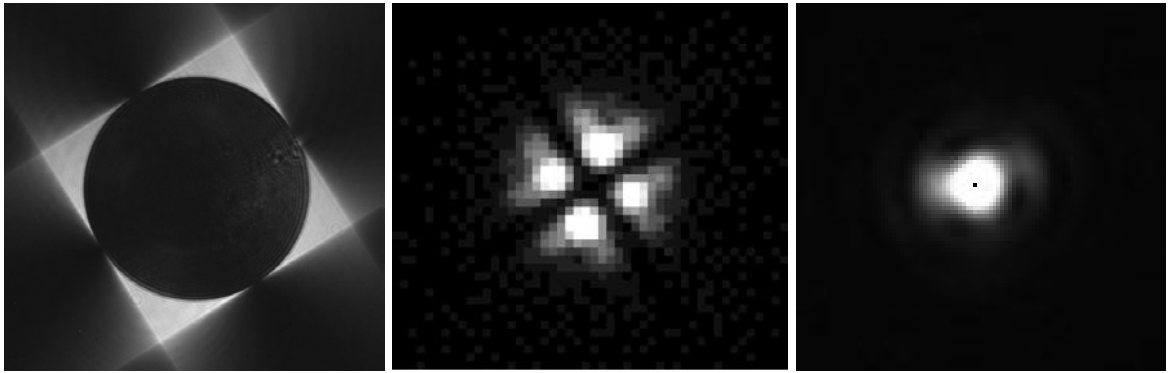

Figure 5. From the left: the image of the FQPM in pupil plane, the FQPM PSF w/o and w the LS.

## 2.3 Raw contrast

The efficiency of a coronagraph resides in its capability to suppress the incoming light selectively with respect to the position of the source in the field: it has to block on-axis light while transmitting off-axis radiation as much as possible. Raw contrast is the ratio between these two components. Hereafter, we will use the terms "off-axis PSF" and "coronagraphic PSF" to refer to:

- Off-axis PSF: the PSF of an off-axis source: it is obtained with the apodizer in the optical path, but no FPM. Here "off-axis" has to be intended inside the discovery region of the coronagraph.
- Coronagraphic PSF: the PSF of a source on-axis: it is obtained using the combination apodizer+FPM.

Raw contrast is defined as:

$$C(r) = \frac{I(r)}{I_0 \cdot M(r) \cdot t_r} \quad (1)$$

Where:

- $I_0$ is the peak intensity of the off-axis PSF.
- M(r) is the FPM intensity transmission curve. For both SPC and FQPM, M is assumed to be 1 in the discovery space and 0 elsewhere.
- $t_r$ is the ratio of exposure times (coronagraphic to off-axis)
- I(r) is the intensity in the coronagraphic PSF. It is computed by taking the azimuthal average of the counts at steps of 6.45um (the pixel size of the camera). The center used for the calculation is the one of the FPM.

## 3. RESULTS

Measured raw contrasts for the SPC designs are reported in Figure 6 and 7, where they are compared with simulated ones. For displaying purposes, simulated contrasts are not zeroed outside the transmissive region of the FPM. Inside the transmissive regions, we found a very good match between measurements and simulations for all three designs. For the simulated curve for the SP1 (that is the symmetric SPC) in Figure 6 the level contrast less than $10^{-5}$ near the IWA is a fake effect, the nominal raw contrast for this coronagraph is in fact $10^{-4}$. For both the asymmetric SPC (we presented SP2b as an example of the asymmetric SPC) the measured raw contrasts are consistent with the simulations.

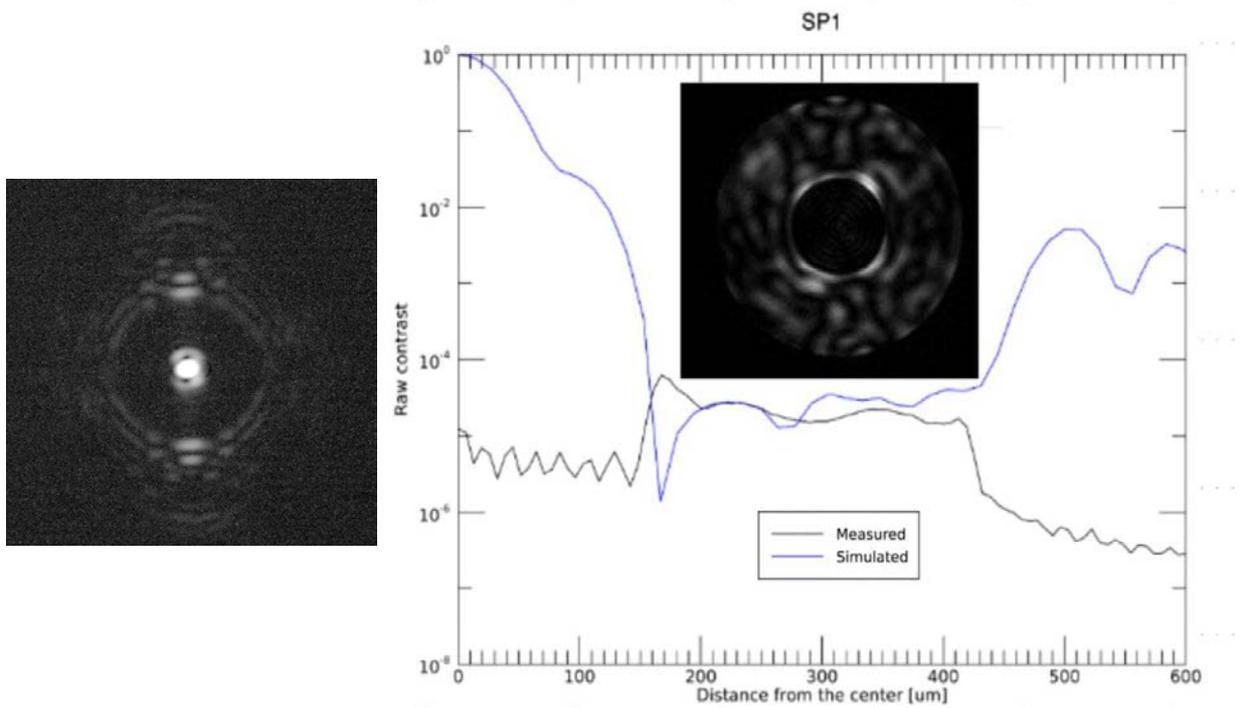

Figure 6. SPC symmetric mask PSF and the achived contrast compared to the simulated one

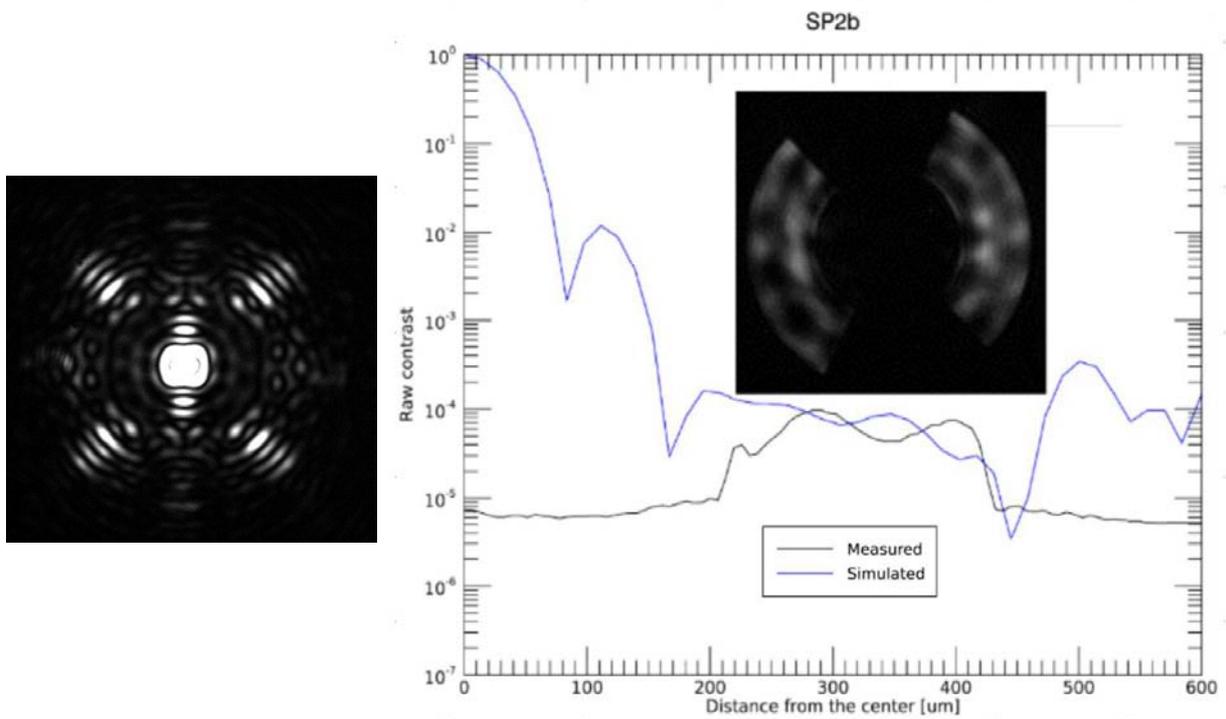

Figure 7. The SPC asymmetric mask PSF and achived contrast compared to the simulated one.

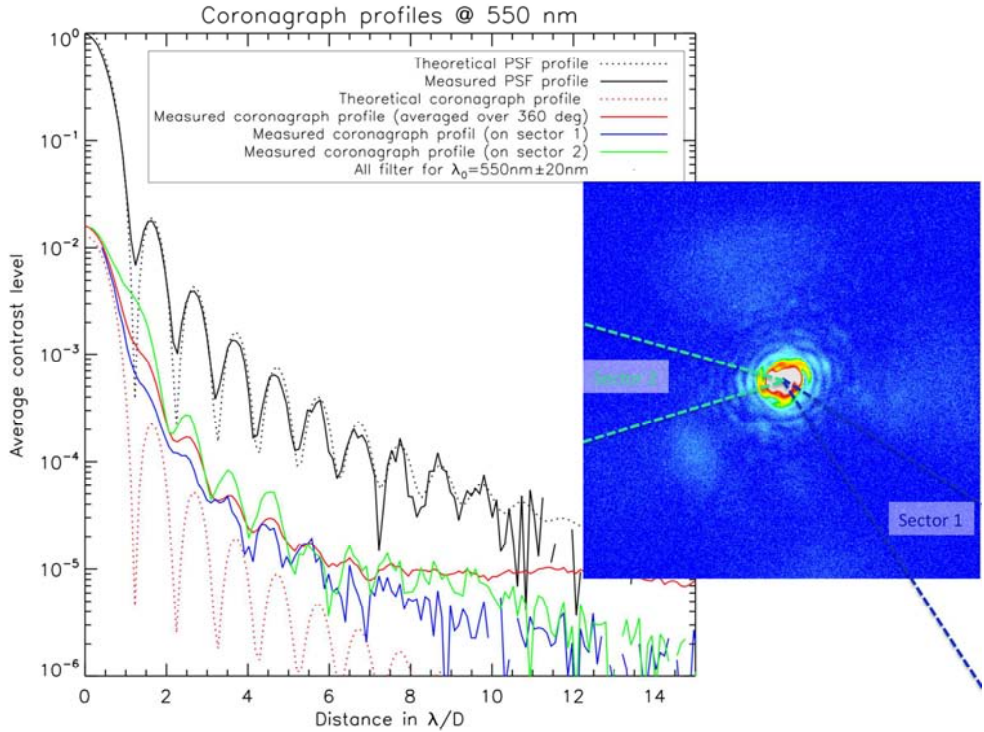

Figure 8. The FQPM achived contrast compared to the theoretical one.

## 4. CONCLUSION

We designed and set up an optical bench to measure the raw contrast for the coronagraphic techniques chosen for SHARK-NIR. The estimated and computed curves are comparable. In view of the SHARK-NIR AIV phase we established the alignment procedures and developed some programs to position the coronagraphic masks with high accuracy.